\documentclass[10pt,conference,final,twocolumn]{IEEEtran}
\usepackage{amsmath}
\usepackage{times}
\usepackage{cases}
\usepackage{amsfonts}
\usepackage{indentfirst}
\usepackage{graphicx}
\usepackage{epsfig}
\usepackage{eucal}
\usepackage{cite}
\usepackage{latexsym}

\usepackage{subfigure}

\IEEEoverridecommandlockouts

\begin{document}

%============
%\frame{
%\begin{minipage}[h]{6.5in}
%\vspace{-0.2cm}

\title{Analog Turbo Codes: Turning Chaos to Reliability\thanks{Li's work is partially supported by National Science Foundation Under Grant No  CCF-0829888, OCI-1122027 and CMMI-0928092.}}
% Tri-State Spectrum Sensing and Distributed Probabilistic Inference for Cognitive Radios}
%\author{\begin{tabular}{c@{\extracolsep{8em}}c}
\author{ Jing Li (Tiffany)$^{\dagger \ddagger}$, \ \  and \ \  Kai Xie$^{\ddagger}$ \\
$\dagger$ School of Electronic Information Engineering, Soochow University, Suzhou, Jiangsu Province, P.R.China \\
$\ddagger$ Dept of Electrical and Computer Engineering, 
Lehigh University, Bethlehem, PA 18015, USA \\
Emails: \ jingli@ece.lehigh.edu,\  kax205@ece.lehigh.edu 
\vspace*{-0.5cm}}
\date{}

\maketitle

\begin{abstract}
%While error correction is generally performed in the digital or the discrete domain, analog or real-numbered error correction codes are also possible.
Analog error correction codes, by relaxing the source space and the codeword space from discrete fields to continuous fields, present a generalization of digital codes. While linear codes are sufficient for digital codes, they are not for analog codes, and hence nonlinear mappings must be employed to fully harness the power of analog codes. This paper demonstrates new ways of building effective (nonlinear) analog codes from a special class of nonlinear, fast-diverging functions known as the chaotic functions. It is shown that the ``butterfly effect'' of the chaotic functions matches elegantly with the distance expansion condition required for error correction, and that the 
useful idea in digital turbo codes can be exploited to construct efficient turbo-like chaotic analog codes. Simulations show that the new analog codes can perform on par with, or better than, their digital counter-parts when transmitting analog sources. 
\end{abstract}
\section{Introduction}
%\label{sec:introduction}
%\vspace{0.3cm}
%\noindent {\bf 1. Introduction and Motivation}
%\vspace{0.3cm}

Seventy years of coding research has resulted in many remarkable efficient error correction codes. The culmination is reflected in the discovery of turbo codes and and the re-discovery of low-density parity-check (LDPC) codes -- practical linear codes that are capable of approaching  the Shannon limit on additive white Gaussian noise (AWGN) channels. Since these codes are digital codes, analog sources, such as sound, color, images, and the vast geo-, physical-, chemical-, bio-signals acquired by the sensors, must first be sampled (discretizing the time) and quantized (discretizing the amplitude). The communication theory states that as long as sampling is performed at or above the Nyquist rate, the discrete-time samples can losslessly recover the original continuous-time signals. In comparison, however, quantization will result in permanent granularity error that is irrecoverable. To keep the granularity error small in general requires more quantization levels and/or higher-dimension quantization (i.e. vector quantization). The former may cause non-negligible bandwidth expansion, and the latter can be very hard to design.    

One possibility to avoid the burden of quantization and the associated granularity error is to transmit the discrete-time continuous-valued analog source directly in its analog form. Since real-world channels are inevitably noisy, to ensure adequate transmission fidelity requires practical analog error correction codes (AECC), or, simply, analog codes -- codes that encode (discrete-time) continuous-valued analog source sequences to (discrete-time) continuous-valued analog codeword to combat channel noise and distortion \cite{bib:Wolf83b,bib:Mars84,bib:vardy,bib:chaotic_analog_coding}. Although not nearly as popular as digital codes, the notion of ``analog error correction coding'' actually dated back to the early 80's, when Wolf \cite{bib:Wolf83b} and Marshall \cite{bib:Mars84} independently introduced the concept. (It was termed {\it real number coding} in  Marshall's work and {\it analog coding} in Wolf's work.)  
%\cite{bib:Mars81,bib:Wolf83a} 
%\cite{bib:Wolf83b,bib:Mars84}. 
To put the digital systems in perspective, the combination of quantization, digital error correction codes and the QAM (quadratic amplitude modulation) digital modulation\footnote{High-order QAM (e.g. 256QAM or 512QAM) has become dominant in high-date-rate wireless systems; and analog data can be naturally modulated through $\infty$-order QAM.} may be viewed as a single analog error correction code. A noteworthy advantage of using a single analog code in lieu of its digital counter-part is the simplicity. To design a good digital system requires not only the careful design of individual components, but also a judicious balance of the rates between them, i.e. how many bits to use for quantization, error correction coding and modulation, respectively. All of this involves a lot of complicated design issues. 

Although analog error correction follows much the same philosophy as digital error correction, good AECCs are hard to find. Early ideas of analog codes were a natural outgrowth of digital codes, by extending  
conventional digital codes from the finite field to the
real-valued or the complex-valued field (namely, symbols from a very large finite field can approximate real values). This has resulted in, for example, analog  BCH codes and analog RS codes %\cite{bib:Mars81,bib:Wolf83a},
\cite{bib:Wolf83b,bib:Mars84}. Since these (linear) analog codes also rely heavily on the conventional decoding algorithms such as the modified Berlekamp-Massey and Forney algorithm, they perform best on a special {\it pulse} channel where noise only occurs to a limited number of coded symbols. For practical AWGN channels, however, the performance of these analog codes can be rather disappointing, since  every coded symbol is distorted by the channel. Xie and Li recently
established a mean square error (MSE) lower bound (i.e. best achievable performance) for linear analog codes \cite{bib:linearCodes}. The fact that nonlinear analog codes can outperform this lower bound clearly speaks for the necessity and benefits of going nonlinear. 

\vspace{0.2cm} 
\noindent \underline{Key Idea Underpinning Chaotic Analog Codes:} 

To design good analog error correction codes, let us re-evaluate the profound idea underpinning error correction, namely, the principle of {\it distance expansion}. 
%In a good digital code, source sequences with small Hamming distance separation are usually mapped to codewords with larger Hamming distance separation. 
%small Hamming distances between source sequences are magnified to larger Hammin 
In digital encoding (a discrete function), the {\it source space} in which elements have
relatively small Hamming distance and may easily get confused
with each other, is mapped to a {\it code space} in
which elements have (much) larger Hamming distance and can
therefore tolerate (much) larger distortion. 
For analog codes to effectively achieve distance expansion and combat distortion would require good continuous functions that can effectively magnify Euclidean distance. For this, we propose to exploit {\it chaotic} functions (or chaotic systems), a specially class of continuous-valued, nonlinear functions with bounded state spaces exhibiting a topological mixing feature. 
Chaotic systems are widely existent in the natural world as well as the engineering world, and many of them can be realized using simple electric circuits. Despite the rich variety of formalities, chaotic
systems share one common property, namely, high sensitivity to the initial
state. Popularly dubbed the {\it butterfly effect,} this property
states that a small perturbation to the initial state(s) of a chaotic
system will cause a huge difference some time later. Although
this butterfly effect is in general viewed as a system penalty, it
can actually be cleverly exploited to satisfy the distance expansion
property required by a good channel code. Specifically, if one
treats the initial state(s) of a chaotic system as the source (to be
encoded), and treats some later states as the codeword (having been
encoded), then the chaotic system naturally enacts a channel encoder
that successfully magnifies the small differences (distance) among
the source sequences to large ones.

This elegant feature was first noted by Chen and Wornell some twelve years ago. In their pioneering work \cite{bib:chaotic_analog_coding}, they proposed the first chaotic analog code, the {\it tent map code}. The first of its kind, this code directly employs a  {\it tent map} -- a
nonlinear, discrete-time, real-valued chaotic map with simple  formulation -- as the chaotic generator to achieve channel encoding. 
A  near maximum likelihood (ML) detector is also developed to perform effective channel decoding \cite{bib:chaotic_analog_decoding}.   
However, in part because the tent map code performs nowhere comparable to digital codes, and in part because chaotic functions are rather foreign to the coding community, the beautiful idea exposed in \cite{bib:chaotic_analog_coding} was largely ignored. 
%slept for a decade before it was
%recently picked up and extensively generalized by the PI and her group \cite{bib:CAT conf}\cite{bib:baker}.

\vspace{0.2cm} 
\noindent \underline{Novel Constructive Mechanism:}

In this paper, we reclaim this intriguing idea and propose a new and effective way to utilize chaotic functions in designing analog codes. Our studies show that direct application of chaotic functions may be insufficient, since a single chaotic function usually has imbalanced protection of some kind, namely, some part of the codeword may be protected (much) weaker and hence are more prone to error than the others. To mitigate this defect and provide balanced protection to all, we propose to exploit the powerful structure of digital turbo codes. Recall that a turbo code is built on two convolutional codes, which are parallelly concatenated in such a way that if one convolutional code produces a low-weight (i.e. weak) codeword, the other will most likely produce a high-weight (i.e. strong) one. As such, the overall codeword weight is rarely very small, thus significantly improving the worst case and reducing the chance for worst case. Borrowing this idea, we can arrange a similar ``buddy system''  by arranging two simple chaotic functions in a parallel concatenation, such that the vulnerable part of one is properly paired with the robust part of the other. Our new codes are to turbo codes, as single chaotic functions (e.g. the tent map code) are to convolutional codes. 

To demonstrate our idea, below we present two constructive examples, using the tent map and the baker's map, respectively. We discuss how turbo-like structures can be devised for component codes to effectively cover for each other, and verify the effectiveness of our approaches through simulations.  We show that the proposed turbo-like chaotic analog codes not only significantly outperform their processor (i.e. the tent map code \cite{bib:chaotic_analog_coding}), but can also perform on par with, or better than, their digital counter-parts (i.e. the combination of quantization, digital coding and digital modulation).

\section{Key Idea of Chaotic Analog Codes and ML Decoding}
\label{sec:ML}

We consider building analog codes from chaotic functions. 
Prominent features of chaos includes continuous but bounded state space, deterministic randomness, nonlinearity, nonperiodicity, topological mixing, and sensitivity to initial conditions. The last is widely known as the butterfly effect, due to Lorentz's 1972 paper, ``Does the flap of a butterfly’s wings in Brazil set off a tornado in Texas?'' This very feature, i.e. the fast-diverging nature of chaotic functions, is usually measured by a  Lyapunov exponent that is $>1$. 

In general, a chaotic function is a real-valued  or complex-valued recursive function in the form of:
\begin{align}
 {\bf x}_i=F({\bf x}_{i-1}),
\end{align}
 where ${\bf x}_0$ is the initial state vector (the seed), and ${\bf x}_i$ is the state vector at time $i$. A straightforward way to build a rate $1/n$ systematic chaotic analog code, such as how the tent map code is constructed, is to feed the source sequence to the chaotic function as the initial state (the systematic part), and to collect $(n\!-\!1)$ subsequent states as parities to protect the source. 

In theory, a code is well defined as long as  the codebook 
is specified. In practice, there is also need for encoding and especially decoding procedure. For digital codes, irrespective of complexity, one can always perform brute-force exhaustive search (e.g. comparing the received sequence with $2^k$ valid codewords) or syndrome decoding (i.e. placing all the $2^n$  sequences in the standard array and performing table look-up) to achieve good performance. Note that such universal procedures become impossible for analog codes, since the codeword space is now continuous, consisting of unaccountably infinite  points all of which are valid. If an analog codes is constructed by directly taking a known chaotic function as the encoder (e.g. the tent map code), then the decoder may employ existing chaotic estimation methods. 
% (such as those based on linear programming, Bayesian networks, and expectation-maximization\cite{bib:dynamic_programming,bib:Baysian,bib:EM}). 
However, as mentioned before, more sophisticated and better-performing code structure would involve the concatenation or compounding of two or multiple chaotic functions, and judicious decoding procedure must be designed case by case.

Consider an analog code $\mathcal{C}$ with mapping $\mathbb{U}^{k} \stackrel{\cal C}{ \rightarrow} \mathbb{X}^{n}$, where the source space $\mathbb{U}^{k}$ 
and the codeword space $\mathbb{X}^{n}$ are assumed to be continuous and differentiable.  
%comprises  a finite number of $t$ subspaces, denoted as $\mathbb{B}_i$ for $0 \le i \le t\!-\!1$, such that the function $\mathcal{C}$ is continuous and differentiable in each subspace. 
Consider transmitting the codeword $\mathbf{x}$ and receiving the sequence $\mathbf{r}$ at the output of a noise channel. We can define the ML decoder for a general analog code as
\begin{equation}\label{equ: ml decoder of analog decoder}
    \tilde{\mathbf{u}}_0^{k-1} =\mbox{arg} \max_{0 \le i \le t-1} (\mbox{arg} \max_{\tilde{\mathbf{u}}_0^{k-1} \in \mathbb{B}_i} \Pr(\mathbf{r}_0^{n-1}|\mathbf{u}_0^{k-1}
    )), 
\end{equation}
where $\mathbf{u}_0^{k-1}$ is short for $(u_0,u_1,\cdots, u_{k-1})$, and 
 $\tilde{\mathbf{u}}_0^{k-1}$ is the decoder estimation for the source vector $\mathbf{u}_0^{k-1}$. 
%Since the function $\mathcal{C}$ is continuous and 
%differentiable in each subspace $\mathbb{B}_i$, where $ 0 \le i le t\!-\!1$, 
Suppose the channel transfer function 
is also differentiable (a condition that is satisfied for most continuous-output  channels such as AWGN channels and fading channels). 
 Suppose there are only a finite number of local maximums for the target function in (\ref{equ: ml decoder of analog decoder}) (again a condition that is generally satisfied for linear and nonlinear mappings), then we will have a finite number of candidates for possible $\mathbf{u}_0^{k-1}$. The ML decoder can compare all of these candidates to identify the best  
$\mathbf{u}_0^{k-1}$ with the largest probability. The complexity of the ML decoder will be linear to the number of candidates (local maximums). 

In general, one may find the local maximum in each subsection by taking a derivative of the target function with respect to $\mathbf{u}$. Since all chaotic functions are by nature nonlinear, to keep down the decoding complexity, we focus on those chaotic functions that are piece-wise linear.

%%%%%%%%%%%%%%%%%%%%%%%%%%%%%%%%%%%%%%%%%%%%%%%%%%%%%%
%%%%%%%%%%%%%%%%%%%%%%%%%%%%%%%%%%%%%%%%%%%%%%%%%%%%%%
\section{Existing Chaotic Analog Codes: Tent Map Codes}
%\label{sec:1-D chaotic analog codes}

%\vspace{0.4cm}
%\noindent{\bf 2. Existing Chaotic Analog Codes: Tent Map Codes}
%\vspace{0.4cm}

The tent map code \cite{bib:chaotic_analog_coding} is constructed by employing a single tent map function as the encoder. The tent map, a simple 1-dimension piece-wise linear function that offers as rich dynamics as infinite length binary shift register, is defined as follows. 
%\vspace{-0.2cm}
\begin{align}\label{equ:tent_mapping} 
& F(x_i) = \beta-1-\beta | x_{i-1} |, \\
\mbox{where \ \ } & 1< \beta \le 2,\ \
 -1 \le x_i \le \beta-1.
\end{align}
Specifically,  parameter $\beta\!=\!2$ is used \cite{bib:chaotic_analog_coding}, such that the tent map maps $[-1,1]$ to itself.  
%Despite the availability of an ML decoder, the performance of the tent map code is rather disappointing, lagging far behind that of the digital codes. 

The tent map code has successfully demonstrated the possibility of
exploiting chaotic systems to achieve error correction, but its performance awaits to be desired. In our study, we performed a careful investigation of the tent map as well as other chaotic functions. Here, coding gain is generally attained through expanding a {\it neighborhood} source (sub)space and hence magnifying the differences (distances) of two close-by symbols. Since the entire space is bounded and each neighborhood subspace gets expanded, two or more neighborhood subspaces that were previously disjoint will have to overlap to sustain the same bounded space. This gives rise to the renowned topologically mixing feature of a chaotic function, at the same time, it also introduces ``backward ambiguity.''   
% and making the very close-by neighbors to separate at an expentially fast speed. 
%%Since the tent map (with $\beta=2$) After n-
That is, a previous state can unequivocally derive a later state (forward determinism), whereas the reverse operation almost always leads to ambiguity (backward ambiguity). Specifically,  there are two values of $x_i$ (same magnitude but opposite signs), both of which can generate $x_{i+1}$. Thus, to deduce $x_i$ from $x_{i+1}$ in the tent map requires the knowledge of the sign of $x_{i}$, denoted as $s_i$. More generally, for parity $x_{n-1}$ to be useful in deriving the source $x_o$, the sign sequence, $s_0, s_1,...,s_{n-2}$ (termed the ``symbolic coding'' in the chaos jargon) must be available.  If the symbolic coding sequence is {\it accurately} known to the decoder, then a parity $x_{n-1}$, which is distorted by AWGN with variance $\sigma^2$, can  guarantee to derive the source $x_0$ with an impressively small mean square error (MSE) of $\sigma^2/2^{n-1}$! 

However, symbolic coding is not known {\it a priori}, and must be estimated from the received symbols. Because of the forward determinism and the backward ambiguity,  state $x_i$ carries information about its own sign and the signs for all the succeeding states (but not the preceding states). That is, the information on symbolic coding is actually  {\it unequally} embedded in the codeword $x_0, x_1, ..., x_n$, with $s_{n}$ being the most reliable (since $x_0, \cdots, x_n$ all bear information of $s_n$) and $s_o$ being the least reliable (only $x_0$ bears information of $s_0$). Such unequal protection is particularly undesirable, because the distortion error introduced by  erroneous $s_0$ is the largest of all -- that is,  exactly where the protection is most needed actually receives the least of it. This probably explains why the tent map code by itself does not perform well. 

% This is in part because it is a one-dimensional code that encodes only a single source symbol $u_i$ at a time.
%From the coding theory, a larger batch involving more source symbols
%at a time can provide a richer context and potentially a stronger
%error correction capability. This suggests that high-dimensional
%chaotic systems may provide better candidates for coding
%construction. Since a high dimension also implies a high complexity,
%below we explore two-dimensional systems to hopefully strike a good
%balance between complexity and performance.
%in a batch  larger coding gain can be %achieved if at longer coding length, because a longer coding length
%leads to more significant time diversity. Therefore, a
%multi-dimensional chaotic analog codes with larger code length is
%desirable if higher error correction capability is pursued.
%To this end, we explore higher-dimensional

\section{Code Design I: Tent Map Turbo Codes}

The previous analysis of the tent map code motivates us to consider a parallel structure that assembles two tent maps in a turbo-like manner for a much needed enhanced protection of the symbolic coding. Recall that a fundamental reason for the remarkable performance of a digital turbo code is that, when one component code produces a low-weight sequence, the other will produce a high-weight one (with a high probability). Recall also that such ``complementary protection'' is achieved by means of interleaving. Exploiting these powerful ideas gives rise to the proposed {\it tent map turbo code}.

%--------------------------------
\begin{figure}[htbs]
\centerline{
\includegraphics[width=3.4in]{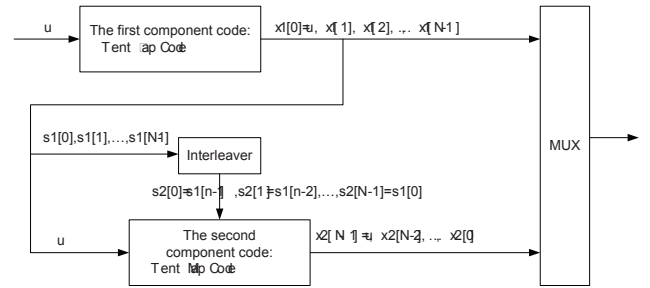}
}
%\vspace{-0.4cm}
\caption{System model for the proposed tent map turbo code.} 
\label{fig:CAT}
\end{figure}
%---------------------------------
\vspace{-0.2cm}

As depicted in Fig. \ref{fig:CAT}, a second tent map is introduced and is assigned a pre-determined symbolic coding sequence: the same symbolic coding sequence generated by the first tent map but in a reverse order. The detailed coding procedure is as follows:

\begin{enumerate}
\item  A source symbol $u$ is used as the initial state and fed to the first tent map (as defined in (\ref{equ:tent_mapping})), generating a (half) codeword $(x_0=u, x_1, ..., x_{n-1})$, which has a sign sequence $(s_0, s_1, ..., s_{n-1}: \ s_i=sign(x_i))$.
 
\item The symbolic coding sequence is scrambled through a ``reverse interleaver'' to get $(s_{n-2}, s_{n-2}, ... s_o)$ (here $s_{n-1}$ is not needed and therefore discarded), and fed to the second component tent map to guide encoding. To make use of the given symbolic coding sequence, the second tent map code is actually encoded backward: using the source symbol $u$ as the last state $x'_{n-1}=u$ and subsequently deriving states $x'_{n-2},...,x'_0$ through the inverse function of the tent map:
\begin{align}
x'_i=\frac{1-x'_{i+1}}{\beta} s_i. 
\end{align}
 %Specifically, in this paper, we let this pre-determined symbolic coding be the reverse oder (reversed indices) of that from the first tent map. 

\item Outputs from both tent maps, $(x_0=u,...,x_{n-1})$ and $(x'_0,...,x'_{n-2},x'_{n-1}=u)$, together form a length-$2n$ codeword, resulting in a code rate $1/(2n)$. Here, the systematic symbol $u$ is transmitted twice, both in the first and in the second tent map. One may also puncture $x'_{n-1}$ and transmit only one copy of the systematize symbol, leading to a code rate of $1/(2n-1)$. 
% may be randomly scrambled before being transmitted. The code rate is $1/(2n-1)$.  
\end{enumerate}

With this coding strategy, $s_i$ with a small indice $i$, which gets 
weak protection from the first tent map, is now gaining 
 a stronger protection from the second tent map. The encoder of this tent map follows the general ML concept discussed in Section \ref{sec:ML}. More detailed discussion, as well as an iterative decoding approach, can also be found in \cite{bib:cat}. 

\begin{figure}[htbf]
%\begin{floatingfigure}{3.3in}
\centerline{
\includegraphics[width=3.4in]{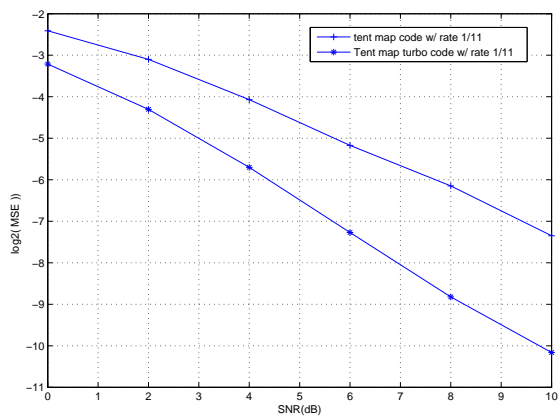}
}
\vspace{-0.4cm}
\caption{Simulation comparison between tent map codes and tent map turbo codes. Code rate: 1/11.}
\label{fig:CATsimu}
%\end{floatingfigure}
\end{figure}

The performance advantage of the resultant tent map turbo code over the single tent map code is verified by computer simulations. Both codes have rate 1/11 and are operated on AWGN channels with ML decoding. Analog source symbols are generated uniformly at random from $[-1,1]$, and the performance is evaluated via MSE (plotted in $log_2$ scale). As shown in Fig. \ref{fig:CATsimu}, the proposed tent map turbo code significantly outperforms the tent map code, with a gain as much as 8 dB. 
%, which confirms the effectiveness of our approach (Fig. \ref{fig:CAT}-right).
%This is close in spirit to that of the classic turbo code, one that tries to make a weakly-protected sequence (i.e. low-weight codeword) from the first component code to hopefully get better protection (i.e. become a high-weight one) from the second component code. 
\vspace{-0.6cm}
\vspace{-0.3cm}

%%%%%%%%%%%%%%%%%%%%%%%%%%%%%%%%%%%%%%%%%%%%%%%%%%%%%

\section{Code Design II: Baker's Map Turbo Codes}
%\subsection{Chen-Wornell Codes}
%\label{sec:coding gain} 

The design philosophy exposed in the previous example has certainly demonstrated an elegant way of exploiting chaotic functions, but there is room for further improvement -- by exploiting more powerful component chaotic functions.
 
The less-than-desirable performance of the tent map code 
may be attributed, in part, to the insufficient and unequal protection for symbolic coding,  and in part, to the low dimensionality of the
underlying chaotic function: The tent map is a 1-dimensional nonlinear
function with a scalar input and offers relatively simple relation
between the time-evolving states. The tent map turbo code 
strengthens the inter-state relation by concatenating two tent maps,
thus creating a higher level of protection, but the input is nevertheless a scalar. From the coding theory, we know that a larger block size will in general offer a richer correlation context and hence promises a better error correction capability. In this second constructive example, we explore a 2-dimensional chaotic function as the component code.  Leveraging the rich literature of the chaos
theory, we identify the {\it baker's map}, a 2-dimensional nonlinear 
function from a unit square to itself, as a desirable candidate. 

\begin{figure}[h]
\centerline{
\includegraphics[width=3.4in]{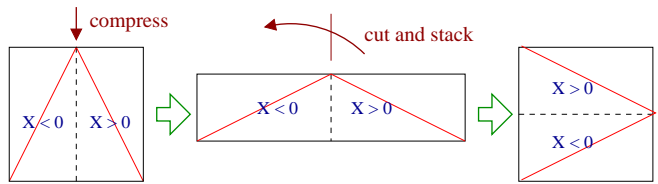}
}
\vspace{-0.1cm} 
\caption{Illustration of the baker's map. }
\label{fig:bakersmap}
\end{figure}

The baker's map is a nonlinear chaotic function named after a 
kneading operation that bakers apply to dough: the dough is compressed and cut in half and the two halves are stacked on one-another. There are two slightly different versions of the baker's map: one may fold over or rotate one of the sliced halves before joining it, or does not fold over the top half. Here we consider the former, i.e., the folded baker's map, which represents a two-dimensional analog of the tent map (see Fig. \ref{fig:bakersmap}. 
Mathematically, the folded baker's map is expressed as: 
%--------------
\begin{align}
 \{x_i,y_i\} &
  =  F(\{x_{i-1},y_{i-1}\}) \nonumber \\
&   = \left\{ \begin{array}{ll}
 \!\!\!\! \{2x_{i-1}\!+\!1, \ \frac{y_{i-1}}{2}\!-\!\frac{1}{2}\}, &\mbox{if $x_{i-1}\!<\!0$} \\
\!\!\!\!  \{1\!-\!2x_{i-1}, \ \frac{1}{2}-\frac{y_{i-1}}{2}\}, &\mbox{if $x_{i-1}\!\ge\!0$}
       \end{array} \right.
        \label{equ: bakers' map of tent map}
\end{align}
%-------------

%--------------------------------
\vspace{-0.2cm}
%\vspace{-0.3cm}
%---------------------------------

Although the baker's map can be directly employed to construct an analog code, just like the tent map code, it does not produce a desirable performance\footnote{A single baker's map performs better than a single tent map, but falls short of the tent map turbo code.}.  
%The baker's map by itself does not benefit analog coding directly,
A close inspection reveals that $F(\{x_{i},y_{i}\})$ is not symmetric, i.e. $y_{i\!+\!1}$ carries information from both $x_{i}$ and $y_i$, but $x_{i+1}$ only carries information from $x_{i}$. 
%again exploit the parallel structure of the digital turbo codes.  
 
\begin{figure}[h]
\centerline{
\includegraphics[width=2.6in]{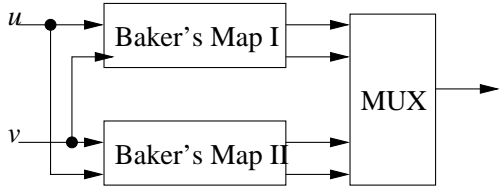}
}
\vspace{-0.3cm} 
\caption{System model of the proposed baker's map turbo code.}
\label{fig:bakers}
\end{figure}

To effectively improve the weak spot and enhance the overall performance, we again exploit the parallel structure of the digital turbo codes.   
%Hence we propose a new coding scheme and make use of
The resultant {\it baker's map turbo code}, as depicted in Fig. \ref{fig:bakers}, comprises  a pair of  baker's maps, engineered in a simple mirrored
replication structure to protect against the weaker dimension of each other.
 Specifically, a pair of source symbols $\{u,v\}$ is fed to the first baker's map as seed $\{x,y\}$, and  fed to the second baker's map as seed $\{y,x\}$.

Similar to the case of the tent map turbo code, the systematic part $u$ and $v$, may be transmitted only once or in both times, resulting in a code rate of $\frac{2}{4n-2}=\frac{1}{2n-1}$ or $\frac{2}{4n}=\frac{1}{2n}$, respectively. 

Note that the baker's map, although being 2-dimensional, is a piece-wise linear function, and since parallel concatenation is also a linear operation, the resultant baker's map turbo code remain a piece-wise linear function. Hence, the same ML decoding philosophy discussed in Section \ref{sec:ML} applies. The entire decoding algorithm is actually quite simple, and the  exact details can be found in \cite{bib:mirroredBaker}.

%--------------------------------
\vspace{-0.1cm}
\vspace{-0.3cm}
%---------------------------------

The simple turbo construction in Fig. \ref{fig:bakers} turns out to be extremely powerful, and the resultant  code demonstrates a surprisingly good performance that is not only better than the tent map turbo codes, but is also comparable to their digital counterparts! 

\begin{figure}[h]
\centerline{
\includegraphics[width=3.4in]{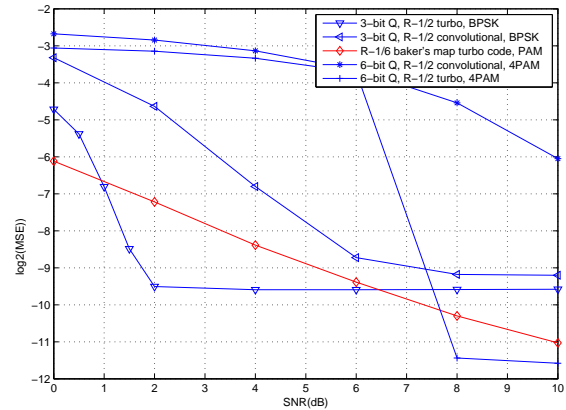}
}
\vspace{-0.5cm} 
\caption{Comparison between the proposed baker's map turbo codes, the digital convolutional coding system, and the digital turbo coding system. All have a bandwidth expansion of 1:6.}
\label{fig:bakersimu}
\end{figure}

Fig. \ref{fig:bakersimu} compares the simulation performance of a digital convolutional code, a digital turbo code, and a rate-1/6 analog baker's map turbo code, in terms of transmitting analog data, uniformly distributed over $[-1,1]$. The convolutional code (8 states) and the digital turbo code (with 8-state recursive systematic convolutional codes as component codes) both have a source block size of 1000 bits, and the baker's map turbo code has a source block size of only 2 symbols (and hence requires much shorter delay and memory size). 
 Uniform scalar quantization (either 3-bit/8-level or 6-bit/64-level quantization) and pulse amplitude modulation (PAM) \footnote{PAM is used, because here the analog code outputs real-valued codewords. Equivalently, one can use QAM for the digital codes, and pack two analog codewords to form one complex-valued sequence  for the analog codes.}  of the appropriate levels are used together with the digital codes, such that the overall bandwidth expansion (i.e. rate of the entire system) is always 1:6. 

Several observations can be made. First, the digital systems is ultimately limited by the quantization error floor (the flat performance curve), whereas the analog code does not seem to have this limitation. Second, a performance trade-off is unavoidable for digital systems. With a high-level quantization (e.g. 6-bit) and hence fewer bits for coding and modulation, the overall performance will have a low quantization error floor, but the waterfall region is also pushed to the far right (the high SNR level). Alternatively, we may allocate more bits to coding and modulation, to push the water-fall region towards the low SNR region, but then fewer bits are available for quantization, which leads to a coarse quantization and hence a high quantization floor. Finally, we see that the baker's map turbo code actually performs comparable to the digital systems -- it consistently outperforms the digital convolutional coding systems, and in some SNR regions also outperforms the digital turbo coding system!

%In addition, tent map codes will suffer an uneven protection for
%different $u$. That is, The distortion is uneven over the variant
%subsection of $u$ for tent map codes. Fig.
%\ref{fig:uneven_protection_over_subsection} shows the difference.
%When SNR is high, like $10$ dB, these $u$ close to the end of $-1$
%and $1$ will obtain a obvious smaller distortion.

%%%%%%%%%%%%%%%%%%%%%%%%%%%%%%%%%%%%%%%%%%%%%%%%%%%%%%
%%%%%%%%%%%%%%%%%%%%%%%%%%%%%%%%%%%%%%%%%%%%%%%%%%%%%%

\section{Conclusion} 

Analog error correction codes, by relaxing the source space and the codeword space from discrete fields to continuous fields, present a generalization of digital codes. 
By cleverly exploiting the ``butterfly effect'' of the chaotic systems, and by designing practical and effective coding structure, we have succeeded in constructing two classes of turbo-like chaotic analog codes: the tent map turbo codes, and the baker's map turbo codes.  
%fully developed three classes of cha nalog coding and its
%decoding algorithm. We proposed the baker's coding, especially we
%initiated the approach of transforming a one-dimensional analog
%coding into a two-dimensional one. Although seemingly simple, this
%approach is demonstrated very effective in performance through
%simulations. Besides, this approach is readily applicable to many
%one-dimensional analog coding schemes. At the receiver side, we
%investigated the ML decoding. Simplified ML decoding is performed
%and the new baker's code  
The fundamental idea underpinning the parallel concatenation is presented, and the general principle of maximum likelihood decoding is discussed. 
Computer simulations show that our new codes outperform the existing chaotic analog codes, and some are even comparable to the conventional digital systems (turbo or convolutional codes)! We conclude by
advocating turbo-like (higher-dimensional) analog coding as a new way to encode and protect analog sources. The analog coding approach is simple, and particularly suitable for channels that are highly varying, where it is difficult to design or adapt to an appropriate quantization/digital-coding/digital-modulation scheme.

%We have studied chaotic analog coding. Starting from Chen-Wornell codes, we investigated the source of coding gain and analyzed the performance pitfalls. We show that the backward decoding algorithm, is close but not exactly ML decoding. Realizing the importance of symbolic coding and the fact that a single chaotic system falls short of protection it, we  we propose a turbo structure to bring in the much desired protection. Soft-iterative decoding based on  SISO MAP algorithms is performed, and the new CAT code outperforms the existing chaotic analog code as well as some digital systems.
% We conclude by advocating analog coding as a refreshing
% way to encode and protect information, especially when the source of information is analog in nature.
% and for each
%chaotic analog code, and described the corresponding iterative
%decoding procedure. Simulation results showed that the proposed CAT
%codes outperform Chen-Wornell codes at the same code rate with 4dB
%gain. Additionally, CAT codes promise better performance than some
%conventional digital communication schemes, such as BPSK modulation
%and repetition codes.

%%%%%%%%%%%%%%%%%%%%%%%%%%%%%%%%%%%%%%%%%%%%%%%%%%%%%%

%\vspace{-0.3cm}

%--------------------------------
\vspace{-0.2cm}
\vspace{-0.3cm}
%---------------------------------

\end{document}